\documentclass[12pt, preprint]{aastex}

\def\<#1>{\langle#1\rangle}
\def\pcite#1{[ref]}

\slugcomment{New manuscript: \today}
 
\shortauthors{Dalia Chakrabarty}
\shorttitle{Looking at the Center of M15, Inversely}
 
\begin{document}

\title{An Inverse Look at the Center of M15}

\author{Dalia Chakrabarty}
\affil{School of Physics \& Astronomy\\ 
University of Nottingham\\
Nottingham NG7 2RD\\
U.K.\\
email: dalia.chakrabarty@nottingham.ac.uk}

\begin{abstract}
  The observed radial velocities and transverse velocities of
  individual stars in the globular cluster M15 are implemented as
  inputs to a fully non-parametric code (CHASSIS) in order to estimate
  the equilibrium stellar distribution function and the
  three-dimensional mass density profile, from which are estimated the
  mass-to-light ratio, enclosed mass and velocity dispersion profiles.
  In particular, the paper explores the possibility of the existence
  of a central black hole in M15 via several runs that utilize the
  radial velocity data set which offers kinematic measurements closer
  to the centre of the cluster than the proper motion data. These runs
  are distinguished from each other in the choice of the initial seed
  for the cluster characteristics; however, the profiles identified by
  the algorithm at the end of each run concur with each other, within
  error bars, thus confirming the robustness of CHASSIS. The recovered
  density profiles are noted to exhibit unequivocal flattening, inner
  to about 0.0525pc.  Also, the enclosed mass profile is very close to
  being a power-law function of radius inside 0.1pc and is not
  horizontal. Simplistically speaking, these trends negate the
  possibility of the central mass to be concentrated in a black hole,
  the lower bound on the radius of the sphere of influence of which
  would be $\gtrsim$0.041pc, had it existed. However, proper analysis
  suggests that the mass enclosed within the inner 0.01pc could be in
  the form of a black hole of mass $\sim{10^3}$M$_{\odot}$, under
  two different scenarios, which are discussed.  The line-of-sight
  velocity dispersion is visually found to be very similar to the
  observed dispersion profile. The enclosed mass and velocity
  dispersion profiles calculated from runs done with the proper motion
  data are found to be consistent with the profiles obtained with the
  radial velocity data.

\end{abstract}


\section{Introduction}
\label{sec:intro}
\noindent
Intermediate Mass Black Holes (IMBH) can serve as vincula to bridge
the discontinuity that is otherwise apparent between the extremities
of the black hole mass distribution. Lately, observational data,
indicating the possibile existence of IMBHs, has been accruing; some of
the recent analyses are due to \citet{simonnatur}, \citet{roberts04},
\citet{terashima04} and \citet{cropper04}, among others. A
comprehensive review of the subject of IMBH, including similar pieces
of observational evidence, is presented in \citet{miller04}. The case
for these intermediate mass black holes has also been buttressed by
the predictions of a black hole of mass of the order of
10$^3$M$_{\odot}$ in the center of the globular cluster M15 by
\citet{gebhardt97} and \citet{gressen02}. However, the analysis by
\citet{simon03} implies a model for the central structure of M15 that
does not necessarily require a central black hole.

The paradigm of globular cluster evolution that we are aware of, from
works by \citet{spitzer87} and \citet{meylan97}, is indeed challenged
if an IMBH is needed to model a globular cluster such as M15.
According to conventional understanding, in a globular cluster with
few primordial binaries, core collapse is stalled by the formation of
binaries via three-body processes. By the time core collapse is
halted, the central density in the cluster is quite high, but this
happens on time scales of the order of a few gigayears, (many times
the half-mass relaxation time). In other words, such clusters spend a
long time with cores in their centers. Runaway growth of a massive
object can occur via collisional processes, and result in an IMBH, if
circumstances are right. \citet{simon02} suggest that runaway growth
is possible in clusters with initial half mass relaxation times less
than 25Myr; they also indicate that the same for M15 is probably
higher than this figure. Moreover, on the basis of their modelling,
\citet{simon02} also suggest that star clusters older than about 5Myrs
and current half-mass relaxation times $\lesssim$100Myrs are expected
to contain an IMBH. The current half mass relaxation time of M15 is
about 2.24Gyrs \citep{harris96}, which is nearly 25 times higher than
this limiting value of 100Myrs. Thus, it is unlikely that an IMBH has
formed in the center of M15 by a runaway process. An alternative
mechanism that could potentially form IMBHs in the centers of dense
stellar clusters was suggested by \citet{miller02}. In this picture,
an initial seed is contemplated to undergo a gradual collisional
growth to transform into a massive object.

Thus, if claims of an IMBH in M15 are true, then we can ratiocinate
that a fresh approach needs to be invoked in order to understand
globular cluster evolution. In light of such profound implications of
these predictions, it is important to make an independent evaluation
of the central structure of M15. This is attempted in this paper. The
algorithm used in this exercise is a fully non-parametric code
(CHASSIS) that was used to constrain the mass of the central black
hole in our Galaxy \citep{dalpras} and was calibrated against N-body
realizations of the Hyades \citep{simonH} and Arches \citep{simonA}
clusters \citep{dalzwart}. The input data for this algorithm is taken
from the observed radial velocities of 64 stars in M15
\citep{gressen02} and the observed proper motions of 1764 stars in
this globular cluster \citep{mcnamara03}.

This introductory section is followed by a brief mention of the
salient points of the data sets that have been used as inputs to the
algorithm. In Section~\ref{sec:algorithm}, the basic scheme of CHASSIS
is briefly discussed. A desription of the procedure followed in this
paper, to calculate the mass-to-light ratio, has been discussed in
Section~\ref{sec:ml}. The results obtained by the algorithm are
presented in Section~\ref{sec:results}. This is followed by
Section~\ref{sec:discussions}. Finally, the paper is rounded up by a
conclusive section.

\section{Data}
\label{sec:data}
\noindent
\citet{gressen02} present radial velocities of 64 stars obtained by
the Space Telescope Imaging Spectrograph (STIS) on HST. This sample of
stars is characterized by very high spatial resolution; the smallest
radius at which kinematic information is available in the sample
($r_{\rm min}$) is only about 0.01pc. This data set is shown in
Figure~\ref{fig:data}. \citet{gressen02} surmised that two of the
stars in this sample were RR Lyrae variables which can have velocity
variations characterized by amplitudes of the order of 50kms$^{-1}$.
Therefore, this pair of stars was ignored when the input data was
ported off to the algorithm.

\citet{mcnamara03} present the internal proper motions of 1764 stars
in M15. This data was taken with the {\it HST} Wide Field Planetary
Camera 2 (WFPC2). The observed proper motions are transformed into
transverse velocities, using the distance $D$ to M15 of 10kpc, which
was the value adopted by the observers and is compatible with the
value (of 9.98$\pm$0.47) according to \citet{mcnamara04}. The set of
transverse velocities is plotted against the observed apparent
positions in Figure~\ref{fig:data}. This data set is taken over a much
larger radial range (from 0.55pc to 2.28pc, approximately) though
$r_{\rm min}$ is nearly 55 times larger in this data set than in the
radial velocity sample. The transverse velocities ($v_{\mu}$) are
calculated as the quadrature of the proper motions $\mu_x$ and
$\mu_y$, scaled to units of kms$^{-1}$.
\begin{figure}
\plotone{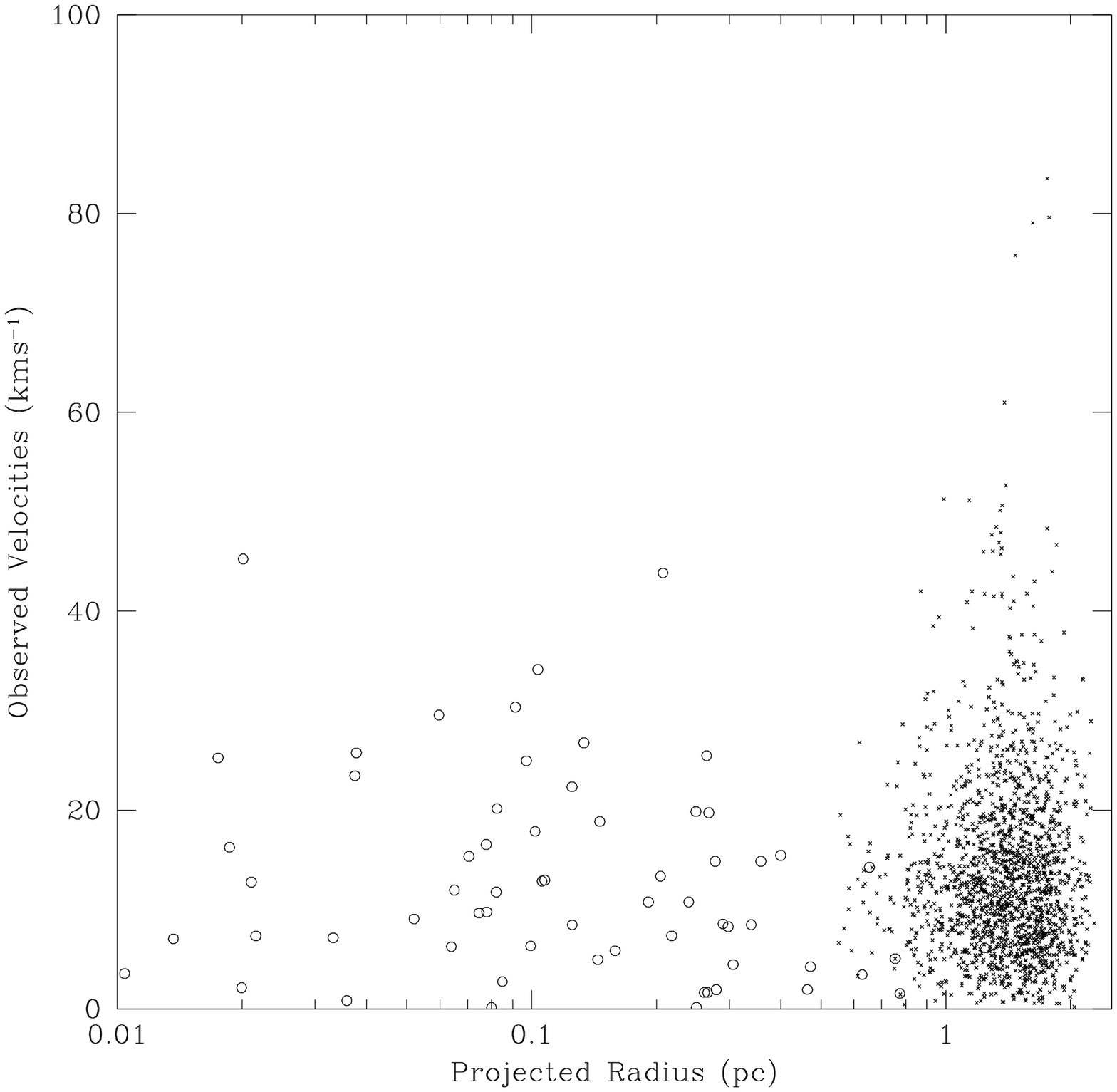}
\end{figure}
Figure~\ref{fig:data} represents the observed kinematic data in a plot
against apparent position. The open circles symbolize the STIS data
wile the transverse velocities are shown by dots. The absolute of
the difference between a radial velocity value and the mean of the
radial velocities is depicted while the transverse velocities are
scaled down by a factor of $\sqrt{2}$ since two components of the
velocity vector define them.

\section{Algorithm}
\label{sec:algorithm}
\noindent
As mentioned in Section~\ref{sec:intro}, the algorithm CHASSIS
(CHAracterizing Stellar Systems with a new Inverse Scheme) has been
used before \citep{dalpras, dalzwart}. It is a fully non-parametric
code that attempts to simultaneously identify the pair of phase
space distribution function (DF) and the potential, (rather the
density from which the potential is subsequently calculated) that best
describe the stellar system at hand. This it does via a maximum
likelihood approach. In its current version, CHASSIS recovers the
system characteristics under the assumptions of sphericity and
isotropy in velocity space. Actually, core collapse clusters are prone
to becoming radially anisotropic in the central regions
\citep{baumgardtmn03}, but this anisotropy is perceived to be small
in the case of M15, at least in the radial ranges that are probed in
this work. Isotropy is less bad an approximation for the runs that
are performed with the radial velocity data than the proper motions,
since the latter data set extends to higher radii. The maxima of
the likelihood function is identified by the Metropolis algorithm.

The algorithm starts with an arbitrary pair of functions that serve as
guesses for the DF and the stellar density. Potential is calculated
from this density, using Poisson equation, under the assumption of
sphericity. At each step, the DF is projected into the space of the
observables at the current form of the potential. This produces the
projected distribution function, which is rendered a function of the
apparent position and the observed velocity component(s). The product
of all such projected distribution functions, each corresponding to a
data point in the observed data set, is the likelihood function. When
the likelihood is maximum, the current choice of DF and density are
said to be the functions that describe the observed data set the best.

The advantage of utilizing the Metropolis algorithm over other
optimization techniques, (such as simulated annealing), is that
Metropolis directly recovers the errors on the DF and density 
from the code; it offers a set of models as possible answers. These
models are distributed according to their likelihoods. The models
lying within the 16$^{th}$ and the 84$^{th}$ centiles are used to
estimate the errors of the analysis. The error bars on the recovered
DF and density are these 68$\%$ errors. All the other quantities of
interest (such as enclosed mass and velocity dispersion profiles)
which are calculated from the density and the DF, are presented with
superimposed error bars that correspond to these errors in the density
and DF. It may be noted that these errors stem from the uncertainties
in identifying the global maxima in the likelihood function and are
essentially different from the observational errors. The errors in the
velocity measurements are incorporated into the analysis by convolving
the projected distribution function with the distribution of the
observational errors, (assumed Gaussian). For greater details about
CHASSIS see \citet{dalpras}.

\section{Mass-to-Light Ratio}
\label{sec:ml}
\noindent
One interesting quantity to estimate from the density distribution
recovered by CHASSIS and the photometry of M15, is the mass-to-light
ratio. The three dimensional mass density identified by CHASSIS can be
projected along the line-of-sight to give the surface mass density
profile of M15. Now, \citet{sosin} provide the surface brightness of
M15 from deep ${\it HST}$ imaging of M15.  The data in \citet{sosin}
is in the form of star counts per square arcsecond while we have the
mass per arcsecond square from the recovered surface mass density
profile.  Therefore, what we need to calculate the mass to light ratio
($\Upsilon$) at any radius, is information on the photometry of M15.
This has been taken from the catalogues presented by \citet{trager},
who provide the surface brightness profile of M15 in the $V$-band.

The quantity that we seek is formally defined as:
\begin{equation}
\Upsilon_X(r) = M(r)/N_X(r),
\label{eqn:m_l}
\end{equation}
where $\Upsilon_X(r)$ is the mass-to-light ratio at radius $r$, in the
waveband $X$. $M(r)$ is the mass of the system in solar masses, at
radius $r$ and $N_X(r)$ is defined as the number of Suns, which when
placed at the distance of M15 (10kpc), would produce the observed flux
at waveband $X$, at radius $r$. Then in the $V$-band, $N_V(r)$ will be
given in terms of the intensity ($\mu_V(r)$) at radius $r$ and
apparent magnitude ($m_{\odot{V}}$) of the Sun at the distance of
10kpc to M15, according to the following prescription:
\begin{equation}
N_V(r) = \displaystyle{10^{\frac{m_{\odot{V}} - \mu_V(r)}{2.5}}}
\end{equation}
Now, $m_{\odot{V}}=M_{\odot{V}} + \log(10\times10^3/10)=19.83$, where
$M_{\odot{V}}$ is the absolute magnitude of the Sun in the $V$-band
(=4.830). The observed surface brightness in the central region is
possibly erronous owing to seeing dificulties. This is averted by
normalizing the surface brightness profile by the value at a radial
location where seeing is least likely to be problem, (namely, the
outermost radius in the star counts data set) using the star counts
data reported by \citet{sosin}.

The errors in the presented mass-to-light profiles correspond to the
1-$\sigma$ errors in the estimation of the surface density, compounded
by the observational errors in the star count profile, obtained from
the data in \citet{sosin}.

\section{Results}
\label{sec:results}
\noindent
A typical feature of all iterative schemes is the requirement of an
initial guess corresponding to the answer that the scheme seeks. If
the algorithm is robust in its design, then the answer will not depend
on the choice of this guess; it is crucial to establish such
robustness on every occasion. CHASSIS has been shown to be robust to
the choice of the initial guesses for the distribution function and
density profile, in its past applications. The independence of the
answers to such choice is discussed below, in reference to runs that
are carried out with the input data set comprised of radial velocity
measurements. Since this data set extends more towards the central
parts of M15 and is better resolved than the proper motion data, it is
more suitable for such analysis.

Now, the form of the density profile that was used as this initial
guess is chosen such that it is is characterized by four parameters: a
scale $\tilde{\rho_0}$, a core radius $r_c$ and two steepness
parameters, $\alpha$ and $\beta$.:
\begin{equation}
\tilde\rho(r) = \displaystyle{\frac
                              {\tilde\rho_0}
                              {\left[
                                1 + \displaystyle{
                                    \left(\frac{r}{r_c}\right)^{\alpha}
                                                 }\right]^{2\beta}}
                             }
\label{eqn:guess_den}
\end{equation}
Assigning values of 2 and 3/4 to $\alpha$ and $\beta$ respectively,
renders the line-of-sight projection of this guess for the initial
form of the density distribution analytical.

At the beginning of the algorithm, the distribution function is
conjectured to be of a form that involves a single free parameter
only:
\begin{equation}
f(\epsilon) = \exp(\epsilon^{\gamma})
\label{eqn:guess_df}
\end{equation}
where $\epsilon$ is the effective energy. The distribution function is
normalized such that its value at the highest effective energy is
unity.

Runs have been performed with assorted values of each of the
parameters present in the functional forms of the initial guesses for
the density and the distribution function. The results from these
various runs have been compared for consistency, to bring forth the
degree of independence of the results from the choice of the initial
guess for the density and distribution function. For easy
interpretation, when one parameter is changed over a wide range of
values, the others are held constant.

\subsection{Effect of changing the initial guess for density}
\noindent
Experiments indicated that the overall scale or amplitude assigned to
the guess for the density distribution does not have any impact on the
results. Even when the amplitude of the initial (guessed) density
profile is varied over 12 orders of magnitude, Metropolis moves ``very
quickly'' to a unique density profile (i.e. the answer) - ``very quickly''
implies within 10$\%$ of the number of steps that are required for
convergence. 

Therefore, the effective number of free parameters in the guess for
the density is three. Three sets of runs were performed to check the
effects of varying these three parameters. Each such set consists of
two runs; the two runs (RUN1 and RUN2) in the first set were marked by
$\beta$ set to 2 and 3, while $r_c$, $\alpha$ and $\gamma$ were held
constant at 0.001pc, 3.8, 1, respectively. Next, $\alpha$ was set to
1.8 (RUN3) for one run and later to 5.8 (RUN4); during both these
runs, $\beta$, $r_c$ and $\gamma$ were set to 1, 0.001pc and 1,
respectively. The third set of runs (RUN5 and RUN6) were carried out
with two varying values of the core radius - $r_c$ was assigned the
values of 0.1pc and 0.00001pc while maintaining $\beta$, $\alpha$ and
$\gamma$ at 3.8, 1, 1, respectively. The setup of each of these runs
is tabulated in Table~\ref{tab:runs}.

\begin{table}
\caption{Description of the eight different runs that are carried out
with the radial velocity data. These runs are characterized by distinct
choices of the initial configurations for the density and distribution
function. These guesses are of the forms represented in
Equations~\ref{eqn:guess_den} and \ref{eqn:guess_df} which involve the
parameters $\alpha$, $\beta$ $\&$ $r_c$ and $\gamma$ respectively. Two
runs are carried out with two widely different values of any one of
these four parameters; while this parameter is being changed, the
other three are maintained constant. Any dependence of the results on
the choice of the initial seed would imply that the algorithm is not
functioning properly.}
\vspace{1cm}
\centerline
{
\begin{tabular}{c|cccc} \tableline
Name & $\alpha$ & $\beta$ & $r_c$(pc) & $\gamma$ \\
RUN1 & 3.8 & 2 & 0.001 & 1\\ 
RUN2 & 3.8 & 3 & 0.001 & 1\\
RUN3 & 1.8 & 1 & 0.001 & 1\\
RUN4 & 5.8 & 1 & 0.001 & 1\\
RUN5 & 3.8 & 1 & 0.1 & 1\\
RUN6 & 3.8 & 1 & 0.00001 & 1\\
RUN7 & 3.8 & 1 & 0.001 & 5\\
RUN8 & 3.8 & 1 & 0.001 & 0.001\\
\tableline
\end{tabular}
}
\label{tab:runs}
\end{table}

The forms of the initial guess in each set of runs and the
corresponding recovered density distributions are shown in the top
panels of Figure~\ref{fig:denml}. Density profiles identified by the
algorithm from runs done with two varying values of the same parameter
are overplotted to aid a visual comparison of the degree of overlap
between the results and the extent by which an achieved profile is
distant from the initial guess for the same. 

\begin{figure}
\plotone{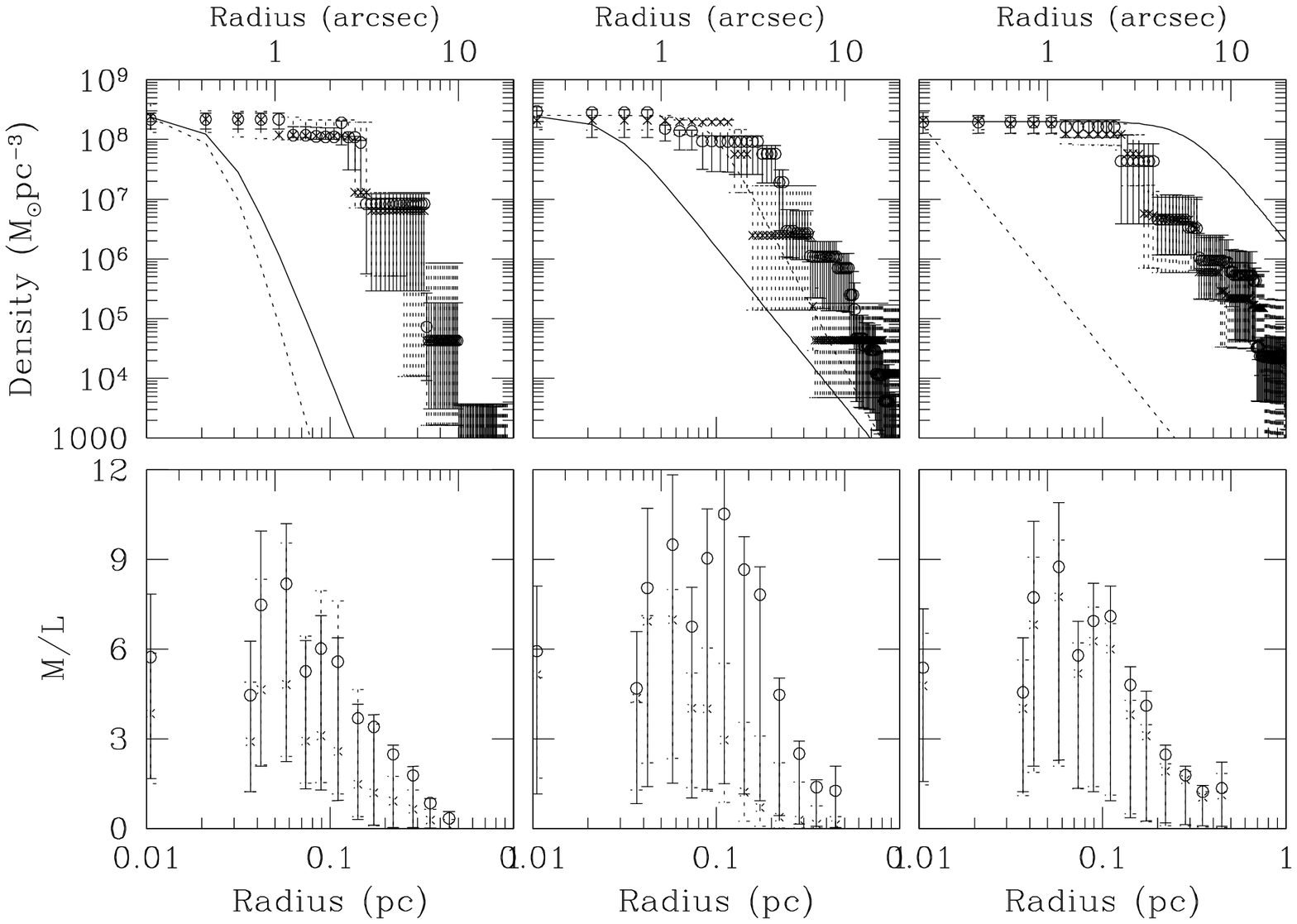}
\end{figure}
\begin{figure}
\plotone{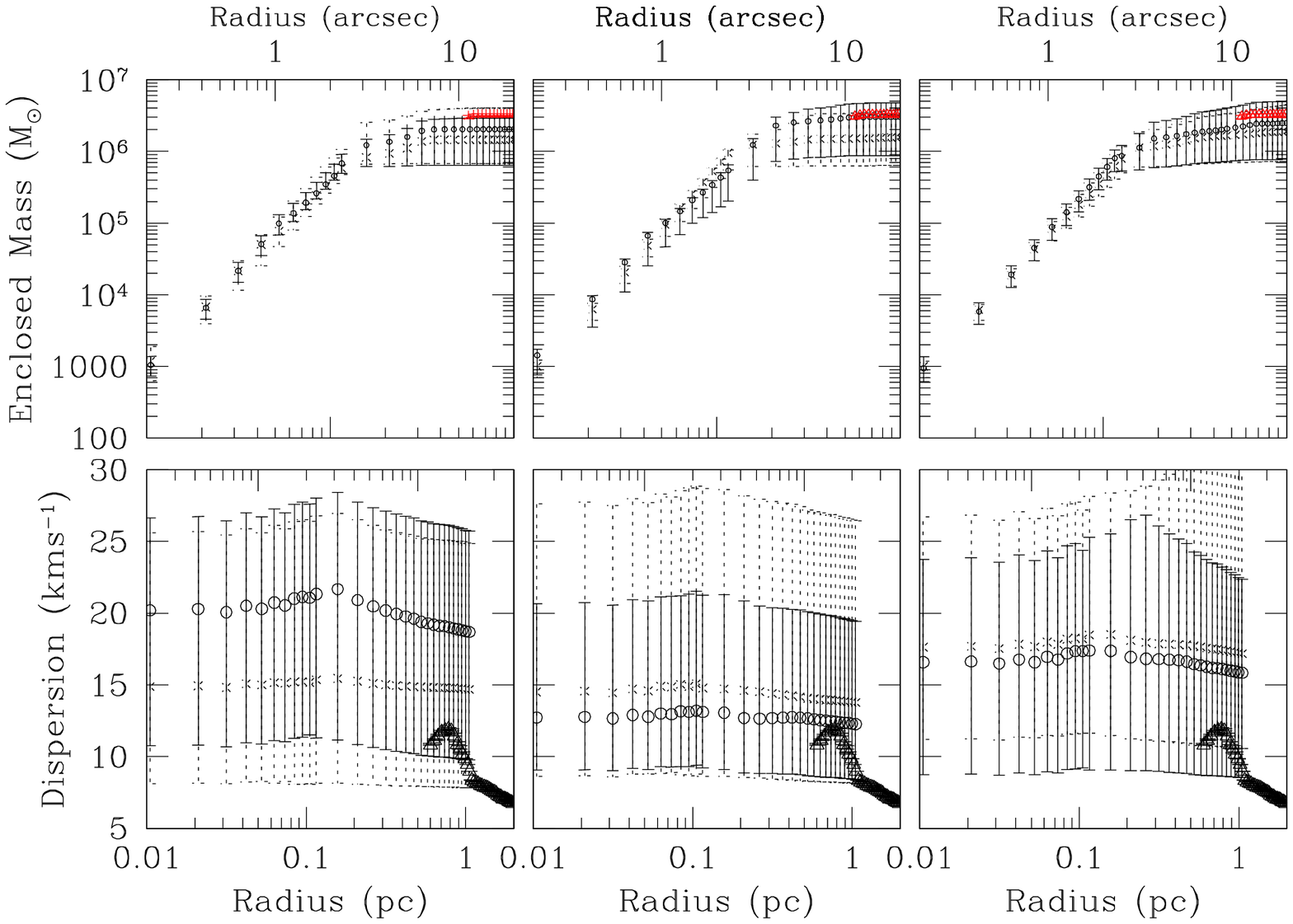}
\end{figure}

As mentioned before, these density profiles are extracted with the
radial velocities that are used as the input to the code. Also, the
recovered profiles are the total (luminous and dark) three dimensional
mass density profile of M15. In each of these runs, the radial
resolution is limited from the bottom by an $r_{\rm min}$ of 0.0104pc
from the observations presented by \citet{gressen02}.  We choose our
radial bins to be 0.0105pc wide, just slightly higher than this
$r_{\rm min}$.

The lower panels in Figure~\ref{fig:denml} display plots of the mass
to light ratio against radius, corresponding to each set of runs that
were performed with the different initial density configurations. The
calculation of the mass-to-light ratio has been discussed before in
Section~\ref{sec:ml}.

Figure~\ref{fig:massvel} shows the enclosed mass and velocity
dispersion profiles obtained from the density distributions and
distribution functions that are recovered by CHASSIS from the
different runs. The mass and dispersion profiles obtained from runs
using the proper motion data are superimposed on these plots.  The
proper motion measurements are characterized by an $r_{\rm min}$ of
about 0.55pc. Thus, the proper motion run was chosen to have a radial
resolution of 0.035pc, with the first radial bin extending from
0.5325pc to 0.5675pc.

An important point to notice is that irrespective of the details of
the run, the recovered density profile is always found to be marked by
a core that extends to about 0.0525pc. Likewise, the enclosed mass
profile is found to be nearly linear in the central parts and is
definitely not horizontal by the first radial bin (0.0105pc). The
mass-to-light ratio profile is found to rise sharply as we go inwards;
this is due to the crowding of non-luminous compact objects in the
central parts of the cluster. This point is taken up for further
discussion in Section~\ref{sec:discussions}. The presented
mass-to-light profiles compare favorably with the same obtained for
the best fitting models in \citet{dull_corr}. Lastly, the velocity
dispersion profiles estimated from the different runs appear to be
consistent with that reported in \citet{gressen02}, upon visual
inspection.

The mass and dispersion profiles found from the proper motion run sit
comfortably within the error bars associated with the same quantities
that are calculated from the radial velocity runs.

\subsection{Effect of changing the initial guess for distribution function}
\noindent
As indicated in Equation~\ref{eqn:guess_df}, the choice for the
initial form of the distribution function is marked by only a single
free parameter, $\gamma$. Two runs are carried out (RUN7 and RUN8),
each with a distinct value of $\gamma$ ($\gamma$=5 and 0.01,
respectively) and the other free parameters assigned constant values,
namely $\alpha=3.8$, $r_c=0.001pc$ and $\beta=1$. Again, these details
are repeated in a tabular form in Table~\ref{tab:runs}.

\begin{figure}
\plotone{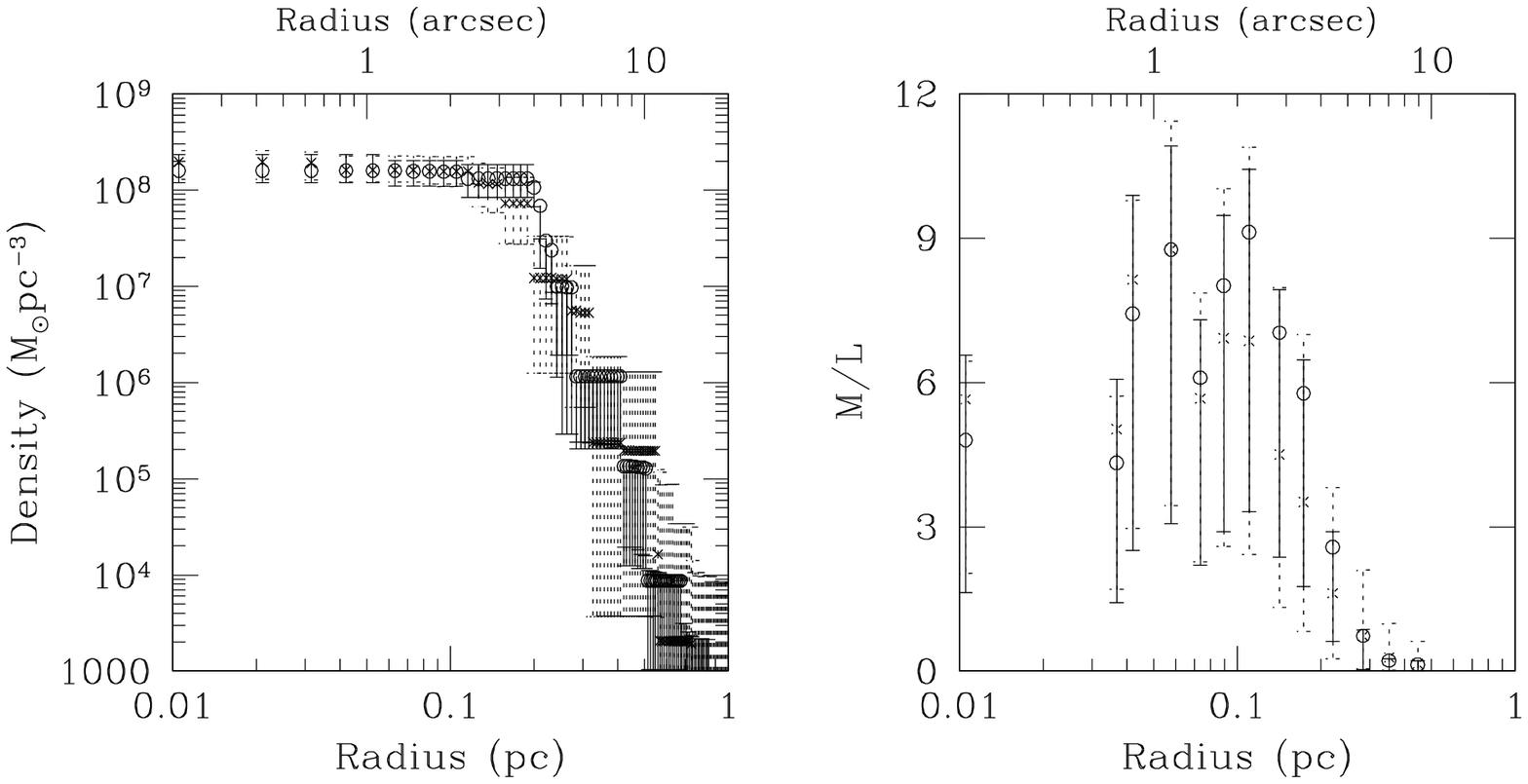}
\end{figure}
\begin{figure}
\plotone{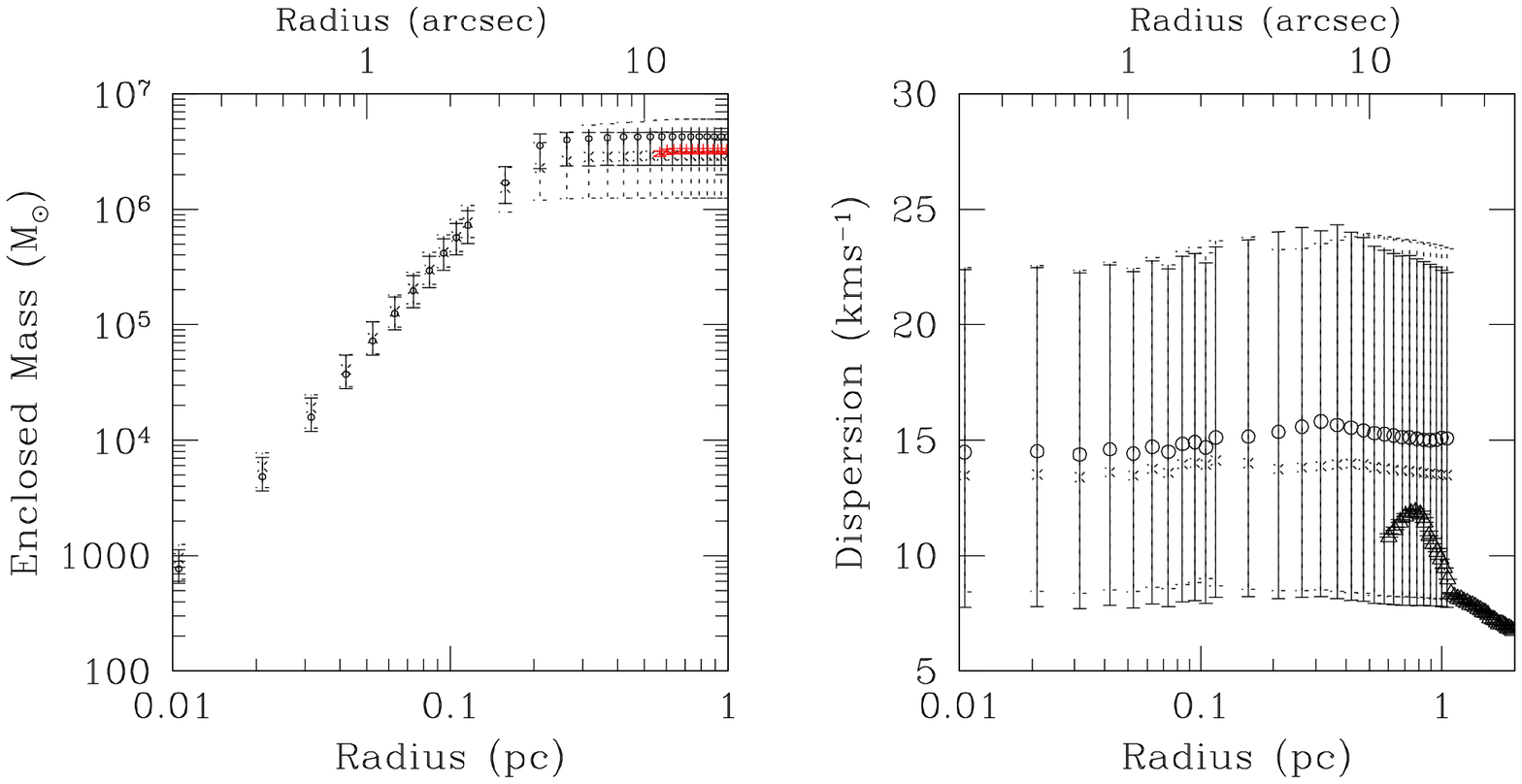}
\end{figure}

The density distributions and mass-to-light ratio profiles obtained
from these runs are shown in Figure~\ref{fig:denml_df}. Again, the
core of size of about 0.0525pc is noted in the density profiles while
the mass-to-light ratio appears to resemble the ones shown in
Figure~\ref{fig:denml}. The enclosed mass and velocity dispersion
profiles, obtained from RUN7 and RUN8 are displayed in
Figure~\ref{fig:massvel_df}. As with the profiles shown in
Figure~\ref{fig:massvel}, we can see that the mass profiles do not
turn horizontal till the right edge of the first radial bin used in
the run.

The results obtained from the different runs (RUN1, $\dots$, RUN8) are
tabulated in Table~\ref{tab:results}. Inspection of the results
obtained from the different runs indicates strong similarities. This
implies that CHASSIS is indeed robust to the selection of the initial
guesses for the quantities that it hopes to identify. This and other
implications of the results are discussed in
Section~\ref{sec:discussions}.

\begin{table}
\caption{Summary of results from the different runs described in
Table~\ref{tab:runs}. All the recovered density profiles are noted to
be cored; the core size forms the second column of the following
table. The mass that is calculated to lie enclosed within the
innermost radial bin is cited as the next column. The projected
velocity dispersion profiles estimated from the distribution function
and density distributions that are recovered from the runs are all
found to be consistent with the observed profile presented in
\citet{gressen02}. Profiles of mass-to-light ratio calculated from the
projection of the recovered density profiles are also noted to be
consistent with the profiles presented in \citet{dull_corr}}
\vspace{1cm}
\centerline
{
\begin{tabular}{c|ccc} \tableline
Name& Size of core of density profile (pc)& Mass within 0.0105pc (M$_\odot$)\\
RUN1 & 0.0525& $1053^{328}_{322}$\\
RUN2 & 0.0420& $1175^{734}_{545}$\\
RUN3 & 0.0420& $1427^{298}_{661}$\\
RUN4 & 0.0525& $1005^{219}_{304}$\\
RUN5 & 0.0525& $948^{412}_{334}$\\
RUN6 & 0.0525& $1005^{176}_{305}$\\
RUN7 & 0.0525& $771^{360}_{192}$\\
RUN8 & 0.0315& $952^{300}_{326}$\\
\tableline
\end{tabular}
}
\label{tab:results}
\end{table}

\section{Discussions}
\label{sec:discussions}

\noindent
In this paper, the observed kinematic data of M15 (radial velocities
by \cite{gressen02} and proper motions by \cite{mcnamara03}) have
been analyzed non-parametrically by the algorithm CHASSIS. The focus
of this work is the very central part of the cluster; the primary
question that this paper attempts to answer is about the existence of a
central IMBH in M15. Since the radial velocity measurements are more
centrally focused than the proper motion observations, they are more
suitable as inputs for the exercise at hand.

Several runs were performed with distinct forms of the seeds for the
quantities that are sought by CHASSIS. Irrespective of what the
initial choices of the distribution function and density were, results
from the runs were found to be consistent with each other, within the
error bars. This establishes the robustness of CHASSIS in the context
of the radial velocity runs and offers confidence in the results
identified by the algorithm.  The recovered density distributions are
always found to exhibit a clear flattening at the center, for
$r\leq{0.0525}$pc in most cases. The core sizes of the profiles
recovered from the different runs are given in
Table~\ref{tab:results}. Since the recovered density profile is the
full three dimensional density distribution, such a flattening is
compelling evidence of the fact that no black hole resides in the
center of M15. 

If a central black hole did indeed exist, then the density profile in
the inner regions of the cluster would have steepened as $r^{-1.75}$,
as the equilibrium stellar density is expected to, in a cluster around
a black hole, \citep{bahcallwolf, baumgardt04a}. The mass profile in these parts would have turned nearly horizontal in this scenario.
Neither profile is noted to behave this way, indicating that the {\it
  enclosed mass is not in a central black hole}. In fact, the enclosed
mass profiles that were inferred on the basis of these recovered
density distributions are noted to vary almost linearly with radius,
inner to about 0.1pc. 

Let us assume, for argument's sake that all the mass that is found to
be enclosed within the innermost radial bin (0.0105pc) is in the form
of a black hole. This quantity, as obtained from the various runs, is
enumerated in Table~\ref{tab:results}. Now, the size of the sphere of
influence of a black hole ($r_s$) could be related to its mass
$M_{BH}$ and the velocity dispersion $\sigma$ by:
\begin{equation}
r_s = \displaystyle{\frac{GM_{BH}}{\sigma^2}}
\label{eqn:r_s}
\end{equation}
where $G$ is the Universal Gravitation Constant. From the enclosed
mass and dispersion data, it can be seen that the minimum that $r_s$
can get is about 0.041pc. The trends in the density and mass profiles
that bear the signatures of a black hole should have been apparent at a
radius that is of the same order as this scale length, if the
assumption that all the mass inner to 0.0105pc was in the form of a
black hole. 

Even under this assumpion, the presented profiles suggest that such
trends are absent at all $r\geq{0.0105}$pc, i.e. by a location that is
nearly 4 times smaller than the $r_s$ of the most massive central
black hole possible from our results. However, it is possible in
principle to arrange the orbits in the central regions of the cluster
in such a way that the effect of a central black hole is limited to
radial locations much inner to this scale length $r_s$, as it is
defined above. Similarly, distributions of orbits can be conjured up
such that the effect of the black hole is felt at locations much
higher than the estimated $r_s$.  Though it may be argued that it is
too contrived to think up orbital distributions that suggest a range
of influence that differs widely from the estimated $r_s$ (in our case
much smaller than the estimated value of $r_s$), it is important to
include such a possibility before arriving at a definite conclusion
about the presence of the IMBH in M15.  In this case, we will not be
able to ascertain the characteristic trends in the density and mass
profiles, unless further data sets at higher resolutions are available.

It is also possible that only a fraction of the mass enclosed within
the right edge of the innermost radial bin is in the form of a central
IMBH. In that case, the sphere of influence of such an IMBH will be
less than 0.041pc and it can be argued that steepening of the density
profile (or flattening of the mass profile) is expected to show up
inner to 0.0105pc. In this scenario, we could miss these trends from
our runs. If this is true, the density distribution of the cluster
will start steepening as $r^{-1.75}$ at some radius $<$0.0105pc, after
remaining flat between about 0.0525pc and this radius. In other words,
the three dimensional density distribution of M15 will be a broken
profile, with the cusp appearing inner to 0.0105pc. Such a cusp can be
betrayed in the results only if the observations improve further in
resolution. 

If either of the two possbilities discussed in the last two paragraphs
is (or both are) true, then under the current state of the
observational data, we can only impose an upper limit on the mass that
may be existent in the form of a central black hole. Values of this
mass are presented in Table~\ref{tab:results} for the different runs;
the results indicate that if the hypothesized IMBH exists, the upper
limit to its mass is $\sim1000$M$_\odot$. If we agree that the radius
of the sphere of influence of a black hole is defined as the location
where the observed central velocity dispersion just balances the
circular speed around the black hole, (as in Equation~\ref{eqn:r_s}),
then to miss the inner cusp in the density distribution, $r_s$ of the
assumed central IMBH has to be $<0.0105$pc. This would then imply that
the upper bound on the mass of such an IMBH is about a quarter of the
mass that we note to be enclosed within our innermost bin, i.e. about
250M$_\odot$.

The velocity dispersion profiles obtained from runs that utilize the
radial velocities in CHASSIS are noted to be similar to the run of the
velocity dispersion reported by \citet{gressen02}. Moreover, the
mass-to-light ratio profiles are found to be consistent with what is
reported in \citet{dull_corr}. Such findings offer confidence in the
results obtained with CHASSIS.

Figure~\ref{fig:massvel} and Figure~\ref{fig:massvel_df} display the
fact that when proper motions are used as the input measurements, the
mass profile from the proper motion data joins in smoothly with that
obtained from the radial velocities, within the error bars.  There are
approximately 30 times more velocity measurements reported in the
proper motion sample than in the STIS data set. This implies that the
errors in the profiles recovered from the proper motion runs should be
about 5.4 (which is $\approx\sqrt{30}$) times less than those derived
from the radial velocity runs. The errors in the innermost bin in the
enclosed mass profile estimated from the proper motion run are about
6$\%$ while these are about 30$\%$ from the radial velocity
runs. Thus, what we note as the extent of the relative errors from the
two types of runs is close to what is expected.

When the proper motions are used as the input to CHASSIS, the
estimated dispersion and enclosed mass profiles are found to be
consistent (within 1$\sigma$ error bars) with the profiles identified
with the radial velocity data as the input. The dispersion results
from the proper motion data do however display the trend of lying on
the lower side of the median level of the mass and dispersion values
from the STIS data. This discrepancy might be an indication of the
anisotropy that is known to affect the velocity distribution in
M15. Also, incorrect estimation of the observational errors could
suggest such a trend.

The run of the recovered mass to light ratio against radius shows a
sharp increase towards the centre though it dips at the centre.  This
increasing trend itself should not be misinterpreted as the signature
of a dark mass in the centre of the cluster. Had the central mass
existed in the form of a black hole, the increasing trend in the $M/L$
profile would have continued unabated to $r=0$. It does not; the trend
indicates the predominance of non-luminous compact remnants such as
neutron stars and white dwarfs in the central parts of this
core-collapse clusters. The effect of the lowering of the fraction of
the less massive visible stars at the very centre is compounded by the
significant reduction in the slope of the density profile as
identified by CHASSIS, as one proceeds inwards.  \citet{simon03} and
\citet{dull_corr} identify a similar result.


The error bars in the recovered density distribution are not
correlated to each other. However, when the enclosed mass profile is
extracted from this density, the errors add up
cumulatively. Consequently, the size of the error bars increases with
radius. Thus, the fractional errors in the innermost bin are about
30$\%$ while at about 1pc the values are much higher. The fractional
errors in the estimation of the line-of-sight dispersion profiles are
expected to be higher than those corresponding to the other derived
quantity, namely the mass profiles. This owes to three numerical
integrals that are involved in the calculation of the density
convolved, projected velocity dispersion profiles, as compared to the
single integral that corresponds to the estimation of the enclosed
mass profiles. Likewise, the errors in the mass-to-light ratio
distributions are expected to be relatively high since the errors in
the estimation of the density by the algorithm are compounded by the
observational errors in the surface brightness profile.

The slope of the recovered mass profile of M15 appears to decrease
significantly by about 1pc. At this radius, the lowest value of the
enclosed mass recovered is about 6.35$\times10^{5}$M$_{\odot}$. Now,
the half-mass radius of M15 is about 3pc, so even if the enclosed mass
at the half-mass radius is still the same as that at 1pc, the mass of
the cluster is at least as high as 1.27$\times10^{6}$M$_{\odot}$. This
value of the predicted cluster mass of M15 appears to be in excess of
what other workers have inferred.  In \citet{dull97} the mass of M15
is cited as about 5$\times10^{5}$M$_{\odot}$ while \citet{mcnamara04}
find it to be about 4.5$\times10^{5}$M$_{\odot}$

A fundamental reason why such a discrepancy exists might be explained
in terms of the assumption of isotropy which is a basic tenet of
CHASSIS. It might indeed be true that M15 is characterized by
anisotropy in the central regions, as \citet{baumgardtmn03} predicted
for core collapsed clusters and \citet{gebhardt00} reported from
ground based observations. The measurement of rotational velocities
were improved upon by \citet{gressen02} - they report a maximum
rotational velocity of 13kms$^{-1}$ at 0.5$^{''}$.
Figure~\ref{fig:excess} describes the situation in light of the
assumption of isotropy in velocity space, even when the system is
really anisotropic. This state of affairs is similar to that of
erroneously assuming an ellipsoidal system to be spherical (see left
panel of Figure~\ref{fig:excess}).  The picture indicated in left
panel of this figure is the same as that described in Figure~12 in
\citet{dalzwart}. The assumption of sphericity leads to an
overestimation of the enclosed mass, i.e. of the quantity given by the
integral $\int{\rho(r)d^3{\bf r}}$, where $\rho(r)$ is the density of
a system (assumed spherical) at the point ${\bf r}$.

Along the same lines, it can be argued that assuming isotropy in
velocity space, when the system is actually marked by anisotropy,
helps to spuriously enhance the quantity $\int{f(E)d^3{\bf v}}$, where
$f(E)$ is the stellar distribution function at energy $E$ which in
turn is given in terms of the potential of the system and the velocity
${\bf v}$. But this integral is really the number density of
stars. Thus, we see that whenever the assumption of isotropy is a
mistake, the number density is artificially increased. Such an
assumption is indeed erroneous near the end of the radial range
considered for our runs done with the STIS data. Hence, we can expect
that at such radii, the number density of stars that is recovered by
CHASSIS is spuriously high. If the mass function of M15 was flat
(which it is not, as reported by Sosin $\&$ vKing 1997), the number
density and mass density would be directly linked. Thus, the
simplistic assumption of isotropy at all radii is expected to lead to
falsely large values of mass densities, at radii around 1pc.

Another way of looking at the same situation is in terms of the
likelihood. The steepness of the distribution function obtained from
the assumption of isotropy is expected to be different from that
calculated under the assumption of anisotropy in velocity space. A
steeper distribution function implies relatively lower values of $f$
at most velocities. This then leads to lower projected distribution
function values and a lower likelihood, during any step. A lower value
of the likelihood function implies a lower binding energy for the most
stable orbit, i.e. less mass enclosed within the orbit. Then the
question is, when is the distribution function steeper; is the profile
of $f$ against velocities steeper when isotropy is assumed or
anisotropy? If the energy dependence of the distribution function is
fixed, the anisotropic distribution function will be steeper.
Therefore, the assumption of isotropy will lead to higher enclosed
mass. Such an assumption appears erroneous at the outer edge of the
considered radial range but not so in the central parts of the
cluster. Hence, it is due to the assumption of isotropy at these outer
parts of our considered radial range, that the enclosed mass will be
rendered higher than if anisotropy were assumed.

Moreover, the lack of sufficient data points around the edge of the
considered radial range could lead to the tail of the recovered mass
profile being less constrained than at inner radii, leading to a
spuriously high cluster mass estimate. There are 37 velocity data
points inner to 0.2pc in the radial velocity data and 25 more data
points that span a much larger radial range from 0.2pc to about 1pc.
In fact, between 0.5pc and 1pc, there are only 4 velocity data
reported in the STIS data.  

Thus, it may be argued that the employment of the ground based data
for M15 would have been a more judicious choice since there are many
more velocity data points in this data set compared to the STIS
observations. However, the ground based observations have much poorer
resolution than the STIS sample.  Therefore, given that the focus of
this paper is the analysis of the ${\it central}$ density structure of
the cluster, particularly the investigation of the existence of the
central IMBH in M15, the STIS data is the pertinent sample.

Thus, we see that the tail of our recovered density profile might be
flatter than what the case is in M15, though the central structure is
correctly predicted. The resulting discrepancy in the recovered
cluster mass owes to the artifacts in the code, discussed above. 

The discrepancy in the cluster mass could also be explained by
carefully examining the reported mass values. In \citet{dull97},
the reported figure is achieved by fitting the Fokker-Planck models of
M15 that the authors have constructed, to the observations of surface
brightness profile, velocity dispersion and the radial profile of the
line-of-sight component of the accelerations of millisecond pulsars
PSR 2127+11A and D. It is to be noted that the corrected version of
the profile of the line-of-sight acceleration has been presented in
\citet{dull_corr}. This work also includes a plot of the velocity
dispersion profiles corresponding to different models (with the
observed data superimposed on top) over an expanded radial range.
Three observations are at least required since there are three
unknowns that need to be effectively constrained to define the best
fit model. These are, the slope of the IMF, the total number of stars
and the radial scale of the model. While the comparison against the
surface brightness profile constrained the good models quite well, the
same could hardly be said about the fit to the velocity dispersion
data. As is apparent from Figure~6 of the paper, or its corresponding
expanded version in \citet{dull_corr}, none of the models fit the
dispersion data well in the range of 0.1pc to 1pc, which is to say
that the good model cannot be distinguished from the unacceptable ones
from this fitting exercise.

However this shortcoming still renders the mass estimate to lie very
close to the prediction by \citet{mcnamara04}, who use the observed
surface brightness data to determine the radial scale by which the
N-body models of \citet{baumNbody} need to be adjusted. The best
fitting model is then found by comparing the observed velocity
proper-motions dispersion profile to the model velocity dispersions,
by a $\chi^2$ test. \citet{gressen03, gressen02} carry out the
modeling of the M15 via the Jeans equation analysis of the observed
radial velocity data (STIS data). Hence it is expected that their
estimate of the total mass is a lower limit to what the true mass is,
given the fact that only a projection of the three dimensional
velocity dispersion profile has been used as the input to the
analysis.

\begin{figure}
\plotone{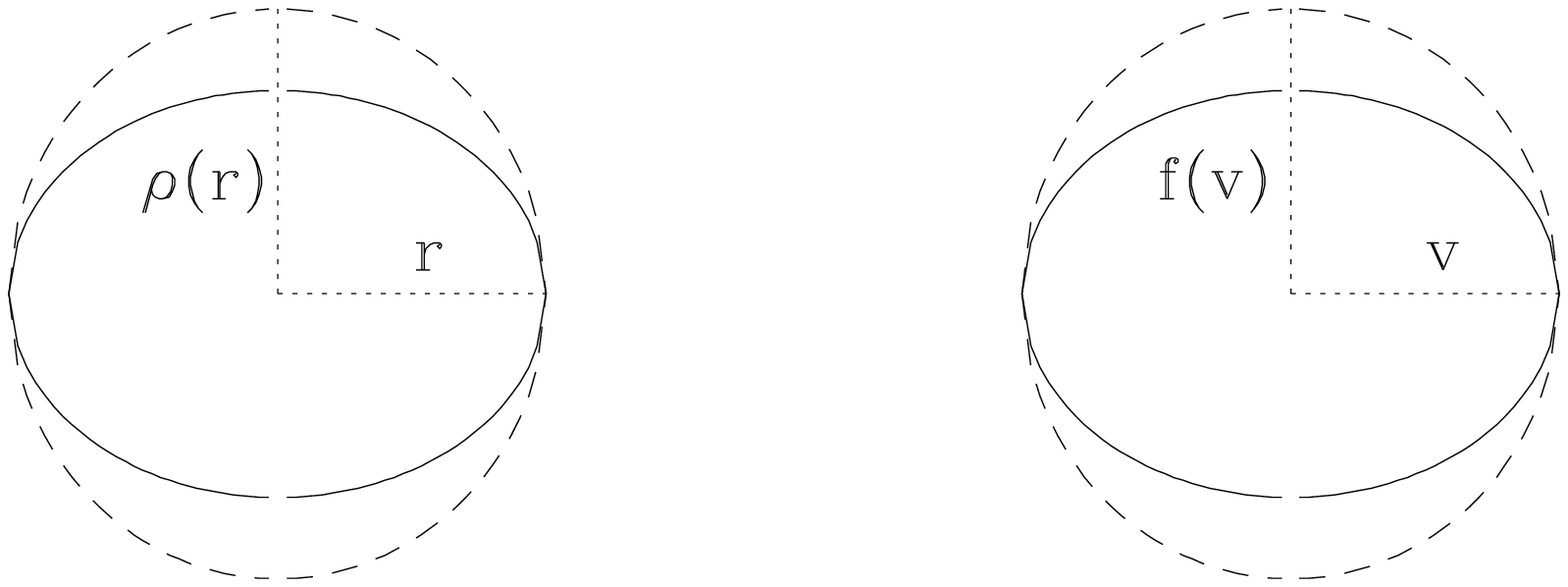}
\end{figure}

It is noted that the density profile suggested by CHASSIS is marked by
jumps, albeit within the errors. The jumps or steps in the recovered
density distribution cannot be physical of course; these are artifacts
typical of any non-parametric algorithm. Traditionally, the way to
smooth out profiles identified by the algorithm, has been by
introducing a bias or a penalty function. However, such a smoothing
function brings owes of its own kind, particularly, the answer then
depends on the choice of the bias function itself. Hence, it is not
the best of choices. Also, smoothing the distribution by hand is
perhaps not an honest reflection of the profile that describes the
data the best. Thus, the kind of profile indicated by CHASSIS is a
better alternative. It is the best description of the observed data
though is rendered ragged by the amplification of the shot noise, a
general problem of the operation of deprojection. On projection of the
density distribution, this problem disappears. The algorithm can
perform better with a larger data set.

In \citet{gressen03}, the earlier analysis done by the same group
\citep{gressen02} on the basis of varying mass-to-light ratios is
amended. Consequently, the newer calculations indicate that models
without a central IMBH are possible within 1$\sigma$ errors. Albeit,
they claim that the fit to the data is ``marginally'' improved by
including a black hole of mass
1.7$^{+2.7}_{-1.7}$)$\times$10$^3$M$_{\odot}$. My analysis however
differs from this inference and my results are closer to the N-body
simulations of the M15 globular cluster by \citet{simon03} who
conclude that the maximum black hole mass is $\leq$500M$_\odot$, in
the extreme case scenario of 0$\%$ neutron star retention from the
cluster at formation (the scenario invoked by \citet{gressen03} to
explain the presence of the central black hole). \citet{mcnamara03}
also claim to find little direct evidence for M15 to contain a black
hole at its center.  \citet{baumgardt04a, baumgardt04b} report the
evolution of clusters that have been assigned central black holes to
start with. These are interesting simulations in which it is studied
how such a cluster would look in projection and how easy it is to
detect the central IMBH in such a cluster. \citet{baumgardt04b} deal
with a realistic mass spectra for the considered clusters; they report
that a density cusp forms around the assigned central black hole, with
the slope of $r^{-1.55}$. Their paper also states that the presence of
IMBHs in ''galactic core collapsed clusters like M15 is ruled
out". This conclusion is compatible with the findings reported in the
current paper, in regard to M15.

\section{Conclusions}
\label{sec:conclu}
\noindent
This paper reports the characteristics of the globular cluster M15, as
estimated by the algorithm CHASSIS. To achieve this, the observed
radial velocities and proper motions of individual stars in M15 are
implemented as inputs to the code. The recovered three dimensional
mass density distribution is found to be distinguished by a core
central to about 0.0525pc. The enclosed mass profile that is
calculated from this density distribution is noted to have the form of
a single power-law in radius, inner to about 0.1pc. These distinct
features appear to be in contradiction with a hypothesis that supports
a central IMBH in M15. This argument appears to be further buttressed
by the following rationale.

The mass that is found to be enclosed within the innermost radial bin
(the extent of which corresponds roughly to the innermost radial
location at which a velocity information is reported) at 0.0105pc is
of the order of 1000M$_\odot$. If we assume that this mass was
entirely in the form of a central black hole, the radius of the sphere
of influence of such a black hole ($r_s$) is found to be at least as
high as 0.041pc. Here, $r_s$ is calculated as the radius at which the
observed (or estimated) central velocity dispersion equals the
circular velocity around the black hole, (see Equation~\ref{eqn:r_s}).
Since this value is nearly 4 times the innermost radius at which the
expected trends in the density and mass profiles are noted to be
absent, the inference is that there exists no IMBH at the M15 centre.
However, such a conclusion can be challenged in the following ways;
firstly the assumption that all the mass is in a black hole maybe
wrong. Secondly, the extent of the influence of a central black hole
is dependant on the phase space distribution and various forms of this
could yield values of $r_s$ that may differ by a factor of few from
the value that Equation~\ref{eqn:r_s} yields.

If only a fraction of the mass that is enclosed within the first
radial bin is in the form of a black hole, then the sphere of
influence of such a black hole would be proportionately smaller than
0.041pc. It is possible that it is even interior to the edge of our
first bin. In this case, we would not be able to spot the expected
trends in our reslts, unless the resolution of the observed kinematic
data sets improved. Of course the density profile then would be
broken; density will rise steeply at some radius $<0.0105$pc, after
staying flat 0.0525pc inwards, approximately. In this scenario, for us
to miss the tell tale signs of the central IMBH, $r_s$ of this black
hole, as given by Equation~\ref{eqn:r_s}, has to be $<0.0105$pc. This
implies that the mass of this IMBH will be less than about a quarter
of the mass that is found enclosed within the first radial bin
(presented in Table~\ref{tab:results}).

The other source of possible concern would be the way the sphere of
influence has been quantified above. It is possible to imagine orbital
distributions around the assumed black hole, which are such that the
influence of the central IMBH is felt at radii much lower than ($<
1/4$ times lower than) the estimated value of 0.041pc. In other words,
it is possible that $r_s$ is not neccessarily the radius where the
central dispersion equals the circular speed around the black hole. In
this case too, the density might steepen up and the mass profile turn
horizontal inner to where the edge of the first radial bin is with the
currently available radial resolution. In this case, we can conclude
that the upper limit on the mass of the hypothesized central IMBH will
be $\sim$1000M$_\odot$.

\section*{Acknowledgement}
\noindent
I am indebted to the following people for their helpful criticisms and
suggestions that contributed towards this paper: Carlton Pryor,
Michael Merrifield, Laura Ferrarese, Simon Portegies Zwart and
Prasenjit Saha. Prasenjit Saha is also acknowledged as the co-author
of CHASSIS. The author was supported by a Royal Society Dorothy
Hodgkin Fellowship.

\clearpage

\figcaption[f1.ps] {Figure displaying the observed data used as
input for the inverse algorithm. The radial velocity measurements of
64 stars in the central core of M15 \citep{gressen03} are plotted as
open circles, against projected radii. What is plotted in the figure
is the absolute of the difference between a measured radial velocity
and the mean of the distribution of the measured radial velocities
(-103.35kms$^{-1}$). The proper motions of 1764 stars in M15 as measured by
\citep{mcnamara03} are transformed into transverse velocities and are
plotted as crosses against projected radii (after scaling the
transverse velocities by $\sqrt{2}$ to account for the fact that two
components of the velocity vector go into defining a transverse
velocity value).
\label{fig:data}}

\figcaption[f2.ps] {The three dimensional (luminous and dark) mass
density (upper panels) and the distribution of the mass-to-light ratio
(lower panels) of M15 as recovered by CHASSIS when radial velocities
of \citet{gressen02} are implemented as inputs to runs RUN1 $\&$ RUN2
(left), RUN3 $\&$ RUN4 (middle) and RUN5 $\&$ RUN6 (right). These runs
are characterized by different choices of the guess for the cluster
density, as parameterized by values of three different parameters
listed in Table~\ref{tab:runs}. The functional forms of the guesses
for the density and distribution function are given in
Equations~\ref{eqn:guess_den} and \ref{eqn:guess_df}. The open circles
superimposed by the error bars in solid lines represent the results
obtained from RUN1, RUN3 and RUN5 while the crosses with the error
bars in broken lines mark the results from RUN2, RUN4 and RUN6. In the
upper panels, the guess for the cluster density profile that the
algorithm starts with is also shown. In RUN1, RUN3 and RUN5, the
initial configuration is represented by the curve in solid line while
the same in RUN2, RUN4 and RUN6 are shown in broken lines. The guess
for the density, as represented here, has been normalized to the value
of the central density that is recovered from the corresponding
run. Density profiles recovered from the different runs are consistent
with each other within the error bars and consistently display a core
which is about 0.0525pc long in most cases. The calculation of the
mass-to-light ratio follows from the discussion in Section~\ref{sec:ml}.
\label{fig:denml}}

\figcaption[f3.ps] {Enclosed mass (top) and line-of-sight velocity
dispersion (bottom) profiles of M15 as estimated from the distribution
function and density that were recovered from runs RUN1 $\&$ RUN2
(left), RUN3 $\&$ RUN4 (middle) and RUN5 $\&$ RUN6 (right). As in
Figure~\ref{fig:denml}, open circles and errors in solid lines
represent the results from RUN1, RUN3 and RUN5 while crosses
superimposed with error bars in broken lines correspond to the
profiles recovered from RUN2, RUN4 and RUN6.  The results from the run
done with the proper motion data is shown in crosses; these profiles
are marked with much smaller error bars than those from the STIS data
runs, owing to the much higher number of velocity measurements in the
transverse velocity data set ($\sim30$ times more). To aid the visual
comparison of the profiles, results from the radial velocity run are
displayed at only few intermediate radii in the range where the radial
velocity and proper motion runs overlap.
\label{fig:massvel}} 

\figcaption[f4.ps] {Total mass density (left) and mass-to-light ratio
of M15 obtained from RUN7 and RUN8. These are runs that are
characterized by a fixed choice of the initial guess for the density
distribution but varying forms of the initial distribution function
that the algorithm iterates upon. These forms are parametrized by the
parameter $\gamma$ that is described in Equation~\ref{eqn:guess_df}
and values of which for these two runs are listed in
Table~\ref{tab:runs}. The density distribution recovered from RUN7 is
shown by open circles and solid lines while the same for RUN8 is shown
in crosses and broken lines. Again, as in the other runs, these
density profiles display a core in the centre and the mass-to-light
profiles are noted to rise quickly at the centre.
\label{fig:denml_df}}

\figcaption[f4.ps] {The enclosed mass profile (left) and projected
velocity dispersion profile (right) as derived from the density and
distribution functions that were recovered from RUN7 (open circles and
solid lines) and RUN8 (crosses and broken lines). The profiles
obtained from the proper motion run are superimposed in crosses on
these distributions. As in Figure~\ref{fig:massvel}, results have been
displayed at intervals, to avoid crowding. The mass and dispersion
profiles shown above are found to be consistent with those displayed
in Figure~\ref{fig:massvel}.
\label{fig:massvel_df}}

\figcaption[f6.ps] {Figure showing the risks of assuming an elliptic
system to be spherical (left) and a system marked by velocity
anisotropy, to be isotopic (right). In the former case, the integral
of the density $\rho$ over all radii $r$ is overestimated by the
simplistic assumption of sphericity, i.e. the enclosed mass is
overestimated at all radii larger than the semi-minor axis of the
ellipse. Similarly, the mistaken assumption of isotropy in velocity
space leads to an overestimation of the integral of the distribution
function $f$ over all velocities $v$, i.e. the number density is
overestimated. This can be a reason why the algorithm predicts larger
densities than is true, at the upper end of the radial scale used in
the radial velocity runs (around 1pc), thereby recovering a cluster
mass in excess of what has been predicted before.
\label{fig:excess}} 

\end{document}